\documentclass[12pt]{article}

\newcommand{\Vec}[1]{\mbox{\boldmath$#1$}}
\newcommand{\Frac}[2]{\mbox{$\displaystyle\frac{#1}{#2}$}}
\newcommand{\Deg}{\mbox{${}^\circ$}}
\newcommand{\Sum}{\displaystyle\sum}
\newcommand{\Int}{\displaystyle\int}

\title{The Thermal Radiation Formula of Planck (1900)}
\author{Luis J. Boya  \\
{\small Departamento de F\'{\i}sica Te\'orica, Facultad de Ciencias
}\\
{\small Universidad de Zaragoza.---  50009 Zaragoza, Spain
}}

\date{luisjo@unizar.es }
\begin{document}
\maketitle


\setcounter{page}{1}

\begin{flushright}
{\sl
This so-called normal energy distribution \\
represents something absolute, and   \\
since the reseach for absolutes \\
has always appeared to me \\
  to be the highest form  \\
of research, I applied \\
myself vigorously \\
to its solution. \\[2ex]
Max PLANCK
             }
\end{flushright}

\label{BoyaP}

\begin{abstract}
We review the derivation of Planck's Radiation Formula on the light of recent
studies in its centenary. We discuss specially the issue of discreteness,
Planck's own opinion on his discovery, and the critical analysis on the
contribution by Ehrenfest, Einstein, Lorentz, etc. We address also the views
of T.S. Kuhn, which conflict with the conventional interpretation that the
discontinuity was already found by Planck.

\end{abstract}

\bigskip\par

\noindent
{\bf 1.-- In the year 2000 we celebrated the 100th anniversary of Planck's
radiation formula}, which opened the scientific world to the quantum. With this
opportunity many papers have appeared, dealing with different aspects of the
formula, its derivation, the meaning for Planck and for other physicists, the
historical context, etc. In this communication we want to recall the origin of
the formula on the light of these contributions, and address some questions,
old and the new, on the meaning of Planck's achievement.

\medskip\par

The fundamental lesson is that the quantum postulate introduces discreteness
and therefore justifies atomicity. Namely the old hypothesis of atoms, started
with the greeks in the Vth century b.C., reinforced through the work on
chemistry in the first part of the XIX century (Proust, Dalton, Prout,
Avogadro), utilized heuristically to explain the properties of gases in the
second part (Clausius, Maxwell, Boltzmann), became a certainty with the
discovery of cathode rays (Pl\"{u}cker), X-rays (R\"ongten), radioactivity
(Becquerel, the Curies), the electron (Thomson) and the nuclear atom
(Rutherford), the later already well within the XX century; in a typical
paradox of science, the same experiments which proved the existence of atoms
also showed, antietymologically, that the atoms were divisible. Now the theory
of quanta has determined the structure of atoms, showing why they do exist in
the first place.

\medskip\par

The historical development of the radiation formula is uncontroversial, and we
shall review it here quickly, referring to the many sources which convey a
detailed information (see the detailed Bibliography at the end). Next we study
the contributions by Planck up to early 1900, when Wien's formula dominated
the scene. The two outstanding communications of Planck to the Berlin Academy
(19-X-1900, guess of the radiation formula, and 14-XII-1900, statistical
justification by introduction of the discrete energy elements) will then ocupy
us. The reception of Planck's discovery was cold, not much being published or
commented until 1905, and we consider why. We refer then to the papers of
Ehrenfest and Einstein, the first to fully realize the big break that Planck's
theory supposes; some contributions by Haas, Sommerfeld and Poincar\'e, and
others are briefly referred. Differences between Planck's energy quanta and
Einstein's {\sl Lichtquanta} are stresssed, in relation to the
indistinguishability issue. We end the historical part by describing
what Planck
did when he came back to the black body radiation problem in 1910--12, the
so-called second theories of Planck.

\medskip\par

We endeavour then to comment briefly on some well-known analysis of Planck's
achievements by Rosenfeld, Klein, Kuhn and Jost. Several centenary
contributions are discussed next, as well as further contributions on the
light of open problems and controversial issues. Our paper ends with a final
look at the figure of Max Planck.

\bigskip\par

{\bf 2.-- That a heated body shines is an elementary observation}; to
understand the dependence of the emitted light (radiation) on the nature and
shape of the body, and on the wavelength and the temperature, is the
{\sl problem of the heat radiation formula}. Experimentally the temperature
ranges were up to 2000 \Deg K, and the vavelength from the near UV up the
medium IR.

\medskip\par

The question was first addressed by Gustav R. KIRCHHOFF in 1859. He discovered
the {\sl universal character} of the radiation law, solving therefore
the problem
of the dependence of the emitted light on the nature, size and shape
of the body;
namely for a ordinary body under illumination there are coefficients of
absorption $a$, and reflection $r$, as well as emission $e$, but the quotient
$e/a = K$, the intensity, Kirchhoff found, is {\sl independent} of the body, if
equilibrium is to be achieved. It depends only on wavelength and temperature;
Kirchhoff hoped the function $K(\nu , T)$ to have a simple form, as is the case
for functions which do not depend on individual properties of bodies. To study
the radiation, one approaches a ``black'' body in which the absorption is
maximal by definition ($a=1$), and studies the radiated intensity as
function of
wavelength and temperature. Empirically it was clear the warmer the body the
greater the total emitted radiation is, and the brightest ``colour'' shifts to
the blue. To realize a black body, one fabricates a hollow cavity ({\sl
Hohlraum}) with the walls blackened by lampblack ({\sl negro de
humo}), practices
a small aperture, heates it up,  and analyzes the outgoing radiation with
bolometers (measure of intensity) and prisms and gratings (measures of
wavelength).

\medskip\par

Since 1865 it was accepted that light was electromagnetic radiation (Maxwell),
and hence the distribution law should be studied by the thermodynamics of the
electromagnetic processes. From the rough experiments performed in the 1870s
it was apparent that the total amount radiated grows like the fourth power of
the absolute temperature, as first stated by J. STEFAN (1879); if
$u d \nu = u(\nu, T)d\nu $
is the differential density of energy of radiation in the hollow cavity
at frequency $\nu$ and temperature $T$,
\begin{equation}
\int_0^\infty  u(\nu, T) d \nu = \sigma T^4 .
\end{equation}

Here the density is $u = 4 \pi K/c$, with $K$ the previous intensity. The
above law
(1) was easy to deduce theoretically (L. BOLTZMANN, 1884). In modern terms, it
follows at once from dimensional analysis with zero photon mass. Next, it was
determined that the wavelength at maximum radiation was inverse with the
temperature ( $\lambda_{\max} T = $ constant, W. WIEN displacement law, 1893);
this is the first example of an adiabatic invariant (Boltzmann), and combined
with the Stefan-Boltzmann's result it yielded  the law
\begin{equation}
u(\nu, T) = \nu^3 f(\nu/T),
\end{equation}
reducing the dependence on frequency and temperature to a single {\sl
universal}
formula on $\nu/T$; notice (2) requires two constants, from
dimensional analysis.
One cannot go any further with pure thermodynamics and the electromagnetic
theory. But by analogy with the velocity distribution formula of Maxwell for
gas molecules, Wien {\sl suggested} the concrete form
\begin{equation}
  u(\nu, T) = a \nu^3 \exp(-b\nu /T)
\end{equation}
which became known as {\sl Wien radiation formula} (1896). The law (3) is very
natural, and indeed for several years it was thought to fit well with
experiments: namely, a power increase (scale invariance) at low $\nu$
followed by
an exponential damping (a cutoff), typical of many physical processes. The
constants $a$ and $b$ should have an universal character, and will play an
important role in Planck's interpretation, see later; they were expected, as
we said.

\medskip\par

However, Wien's law (3) implies that for very high $T$ the density goes to
constant with $T$, which is not very physical: one should expect the energy
density to grow without limit with increasing temperature; this problem does
not arise in the case of Maxwell distribution, which refers to velocities, not
density. Indeed, the refined experiments carried out in Berlin since 1900
mainly by Rubens and Kurlbaum proved that the density, for very low frequency
(equivalent to large $T$), is proportional to $T$.
\medskip\par
For simple derivations of
formulas in this paragraph see the  Appendix. Best secondary sources for this
period are [Born 46], [Jammer 66], [Kuhn 78] and [Sanchez-Ron 01].
\medskip\par

\medskip\par

{\bf 3.-- Now enters Max PLANCK} (*Kiel 1858; \dag G\"ottingen 1947). He was an
expert on thermodynamics, very much impressed by the absolute things, like the
(first) law of conservation of energy, stated along 1840-50 by Joule, Kelvin,
Helmholtz and others, and also by the second law, the increase of entropy,
discovered by his admired R. Clausius first in 1850, later by Kelvin (1853).
At the time, the mechanical theory of heat was accepted, that is, heat is just
another form of energy, not an {\sl entelechia}, the {\sl flogiston};
the idea of
reducing physics to mechanics dominated. For Planck, mechanics represented the
best way to understand physics (and chemistry), and he sought a {\sl
mechanical}
explanation of the second law, understood as an exact law of nature,
on the same
footing as the energy conservation law. By ``mechanical'', Planck did not mean
the atomistic point of view, but continuum mechanics; in fact, for a
long time he
considered the atomic hypothesis something irrelevant (if not nocive) for the
second principle, because in the kinetic theory of gases the second law is not
absolute (Maxwell demon). He stands between Boltzmann, in an extreme, who
always put atomicity first, and the energeticists (Ostwald, etc.), who negated
atoms (as Mach did), pretending to reduce all phenomena to different forms of
energy, on the other. The mechanical model he had in mind was rather close to
the continuum aether of electromagnetism, and Planck sought to prove the
second law from continuum mechanics, in particular the irreversible approach
to equilibrium.

\medskip\par

This is the route which took him to the {\sl W\"armestrahlung}: he thought the
radiant energy of a heated body to be an ideal system to prove the entropy
increase as consequence of {\sl conservative} laws. He also hoped, in the
process, to find ``Kirchhoff function'', that is, the radiation formula. The
heroic struggle of Planck, and his final defeat, is a paradigmatic example of
how an investigation doomed to fail could lead, if pursued
intelligently and with
honesty, to a fundamental, but completely unexpected discovery; by failure it
is meant here that the second law in Planck's form $\Delta S \ge 0$ for $t > 0$
(with $S$ the entropy) could {\sl not} be proven from the blackbody radiation
theory {\sl alone} any more that it could not be proven in the
kinetic theory of
gases without the {\sl Stosszahlansatz} (molecular disorder) of Boltzmann.
\medskip\par
In
five papers, 1897-1899 Planck tried to explain the origin of irreversibility in
the physics of thermal radiation. As the material in the cavity is irrelevant
(Kirchhoff), he considers an oscillator, imitating the resonators
used by Hertz,
constituted by a vibrating dipole $q \Vec{r}$ in presence of an electromagnetic
field ${\cal E}$; the dipole radiates energy at the rate
\begin{equation}
             P = - dE/dt =2q^2/3c^3\,\| \ddot {\Vec{r}}\|^2
\end{equation}
and the differential equation for the dipole amplitude $\Vec{r}$  is
\begin{equation}
         m \ddot {\Vec{r}} + k\Vec{r} - \gamma \ddot {\Vec{v}} = q {\cal E}
\end{equation}
where $m$ is the mass, $k$ the oscillator constant, $q$ the electric charge,
${\cal E} = {\cal E}(t)$ the external electric field, and the cubic term is due
to the radiation damping, with $\gamma = 2q^2/3c^3$. The damping is
conservative,
because energy is not lost to the whole system, it is just transformed into
radiant energy (in contrast to damping by friction, in which case is
transformed
into heat). The damping is small, however, hence in zeroth order
$ \ddot {\Vec{r}} = -(k/m) \Vec{r}$ and Eq. (5) is
\begin{equation}
                K \Vec{f} + (2K/3Lc^3)\dot{\Vec{f}}+ L\ddot{\Vec{f}} =
{\cal E	}
\end{equation}
where $\Vec{f} = q\Vec{r}$ is the dipole moment, $K = k/q^2$ and $L = m/q^2$.
Neither (5) nor (6) by themselves are invariant under time reversal
(because the
odd derivative terms change under $t \to -t$), and this is why and where
Planck hoped to reach irreversibility. Eq. (6) is easy to solve for a given
field ${\cal E	}	$, giving a transient plus a sustained wave
(see Appendix).

\medskip\par

  However, the via towards irreversibility is really closed, as Boltzmann
pointed out inmediately: the whole system of equations for the resonator plus
the e.m. field is time-reversal invariant, and the transformation of the
incoming e.m. plane wave contained in ${\cal E	}	$ into the
scattered (radiated)
outgoing spherical wave does not represent an irreversible change: the theory
allows perfectly well an incoming spherical wave ``scattered'' into
the outgoing
plane wave: grudgingly, Planck had to admit that irreversibility, in the
{\sl W\"armestrahlung} as in the kinetic theory only obtains by explicit
exclusion of some improbable situations: the molecular disorder of
Boltzmann has
here as counterpart the ``natural radiation hypothesis'' of Planck,
imitating the
former. The natural radiation is the one for which there is no correlation
between the phases of the different Fourier components of ${\cal E
	}	$. This is
the defeat of Planck: irreversibility does not come from conservative mechanics
unless extra hypothesis; Planck admitted this around early 1898: the validity
of the second law for the thermal radiation will be, like in Boltzmann's
$H $-theorem, not absolute.

\medskip\par

However, the route to obtain the universal Kirchhoff formula remains open: the
situation at equilibrium is easier to describe that the approach to it. The
next important result of Planck is to admit equilibrium between the resonator
(by now called oscillator) and the bathing radiation, and then to relate the
radiated power (energy per unit time) that is, the loss of oscillator energy,
to energy absorption from the field. At the end, there is a simple relation
between the radiation energy density $u = u(\nu,T)$ and the
elementary oscillator
energy $U$; it is
\begin{equation}
                       u = \frac{8\pi \nu^2}{c^3} \, U,
\end{equation}
a fundamental result in Planck's research. The simplest way to understand (7)
is by dimensional analysis, as $U$ is an energy, but $u$ is an energy density:
$(\nu/c)^2 d(\nu/c)$ is a differential inverse volume; $2 \times 4 \pi$ comes
from polarization and angular integration.

\medskip\par

To solve for the radiation formula, Planck needs to know the mean oscillator
energy $U$ in a thermal bath at $T$. To find it, Planck takes an
indirect route,
retorts to thermodynamics. Instead of guessing $U = U(\nu, T)$, he
did guess the
entropy dependence on $U$, and his Ansatz (obtained by working backwards from
(3)) was a relation between the oscillator entropy $S$ and $U$:
\begin{equation}
   R \equiv - (\partial^2S/\partial U^2)^{-1} = \alpha U
\end{equation}
as the simplest (and, he first thought, unique) possibility. For $\alpha > 0$,
this ensures entropy should increase with time (we do not show here the
arguments, but see later), in agreement with the second law. Together
with $dS =
dU/T$, (8) leads easily to the form
\begin{equation}
    u(\nu, T ) = a \nu^3 \exp(-b \nu/T).
\end{equation}

That is, the law of Wien! We are already in early 1900. Because the universal
character of the Kirchhoff function, the actual formula (9) is universal, and
so are the constants $a$ and $b$; but at the time, two other
universal constants
were known, the velocity of light $c$ and $G$, Newton's constant. So, we have a
perfectly natural system of units, defined by natural primary phenomena, not
anthropomorphically: Planck is exultant and claims that extraterrestian people
would have the same constants, too; for a conservative man like Planck was,
this is an astonishing affirmation!  Later Planck realized that only $a$ is
really new (see later; $b$ should be related to $a$ and to the
constant of gases
$R$ and ultimately to Boltzmann's constant $k$), and this is better:
centimeter,
gram and second, defined humanely, are traded by $G$, $c$ and $a$ (eventually
proportional to $h$). It is remarkable that the universal feature of the
Planck's constant $h$ was realized {\sl before} $h$ was connected with
discontinuity (!), a point insufficiently emphasized by {\sl cognoscenti}; even
[Kuhn 78] to whom we follow in part, does not attach too much significance to
it. The point is, already in the Wien displacement law $\lambda_{\max} T =$
constant, there is a new universal
constant. To complicate matters more, at the time it was not known that $b$
embodies also Boltzmann constant, never written as such by Boltzmann, by the
way.

\medskip\par

\medskip\par

{\bf 4.-- That was nice, as Wien's formula was thought to be correct at the
time}. In fact too nice: as mentioned, the experiments carried out
along 1900 at
the longest infrared waves available, up to 60 $\mu$, showed inequivocally that
$u$ is linear in $T$ for $T$ large, as it was physically reasonable
(see above).
It is astonishing what a simple modification of (8) brings the
experiment data in
order: Planck admits simultaneously that (8) is not unique, and realizes that
in the limit $u \propto T$ one should have (Cfr. (8))  $R \propto U^2$, so the
simplest interpolation formula (with both $\alpha, \beta > 0$)
\begin{equation}
  R = \alpha U + \beta U^2
\end{equation}
leads, by the same steps as before, to the formula
\begin{equation}
    u(\nu, T) = \frac{a \nu^3/c^3}{  \exp(b \nu /T) -1}
    \qquad (\mbox{Planck, 19-X-1900})
\end{equation}
in perfect agreement {\em then and always} with the refined experiments, with
similar two constants as before: this is why we insist that $h$ (or rather
$b$), by
itself, is introduced already at the Wien's formula level. To see the accuracy
of (11) we just recall that the ever-pervading cosmic microwave background
radiation fits to the formula (11) for $T = 2.726$ \Deg K at nearly
the millionth
level precision (COBE device, 1993; WMAP, 2003).

\medskip\par

\medskip\par

``Never in the history of physics was there such an inconspicuous mathematical
interpolation with such far-reaching physical and philosophical consequences''
[Jammer 66, p.18].

\medskip\par

As Pais has remarked [Pais 82, 19a], had Planck stopped with (11), he would
have always be remembered as the man who found the radiation formula, and
would have a place among the greats. The fact that he went on, to supply a
theoretical support for (11) is a measure of his greatness. It represented for
him the biggest effort in his life. He had to yield to Boltzmann again (``an
act of desesperation''), this time, against his most intimate convictions, that
is, to the statistical considerations which have had enabled Boltmann to prove
the $H$-theorem. The reason, or rather the lack of it, is that statistical
considerations were the only ones, known around 1900, to calculate entropies,
by the probabilistic formula chiseled in Boltzmann's tomb in Vienna
$S = k \log W$, with $W$ probability ({\sl Wahrsheinlichkeit}).

\medskip\par

Planck follows a combinatorial method traced from Boltzmann, including the
apparently innocent discretization of energy, $E = \varepsilon, 2\varepsilon ,
3\varepsilon,\ldots$, but the crucial point is that to obtain the
expected result
(11), contrary to Boltzmann, he cannot take $\varepsilon \to 0$, because
then only the $u \propto T$ limit obtains (this again follows today from the
zero mass of the photon). The combinatorics is simple and  the final result is
\begin{equation}
   u(\nu, T)=\frac {8\pi \nu^2}{ c^3} \frac{ h \nu }{\exp (h\nu /kT)-1}
\qquad \mbox{(Planck,
14-XII-1900)}
\end{equation}
with the necessary identification $\varepsilon = h \nu$ to satisfy the
displacement law. We have restored to today's accepted constants $k$
and $h$; it
turns out that $k = R/L$, the ratio of the gas constant to Avogadro-Loschmidt
number. Planck christened $k$ as Boltzmann's constant; he also referred to his
discrete $\varepsilon, 2\varepsilon, \ldots$, as {\sl energy quanta}. Planck
wrote $h$ for {\sl Hilfsgr\"osse}; this is what he first though of it
(I owe this
remark to M. F. Ra\~nada (Zaragoza)).

\medskip\par

This is how the {\sl quantum} entered first time in physics, to
remain for ever,
altough it is true that Planck did not pay, at the time, too much importance
to it. By nearly universal consent, the date December 14, 1900 is considered
the birthday of quantum theory, and a century later it has been duly
celebrated throughout the world.

\medskip\par

\medskip\par

{\bf 5.-- Historically it is very clear that Planck had introduced the finite
energy elements $\varepsilon$ as Boltzmann did}, to calculate entropies from
denumerable entities: he already saw that, contrary to Boltzmann, the size of
the energy element cannot be taken to be zero, and to this he attributed his
discovery. Hence he did realize he had found something important, and
because of
this our point of view is closer to that of Mehra-Rechenberg [Mehra 82] than to
[Kuhn 78], that he indeed realized the new discovery, although did not
perceived, yet, that he had discovered the discontinuity, and in this point we
think Kuhn is right.  He also saw that $h$ had dimensions of action, and the
identification of
$k$ allowed him to compute Avogadro's number $L$ and the elementary charge $e$,
by taking $R$, the gas constant, and $F$, the Faraday, as known at
the time, and
computing $k$ and $h$ from the black-body fit. With the scale molar
vs molecular
settled  (i.e., $L$ known) and also the elementary electric charge, Planck was
to be henceforth a devoted atomist [Heilbron 86, p. 23]. Only E. Mach, among
the notable, remained antiatomist until his very death in 1916.

\medskip\par

The reception of Planck's formula and theory was cold. Out of stressing the
beautiful experimental fit, people were not very keen with the obscure
reasonings of Planck, and the black body physics was a pretty isolated corner
of the general physical research (much centered, at the time, in radioactivity,
the photoeffect and X-rays). One should add that the sheer number of
researchers in physics in the world was then perhaps a hundredth of today's.
Still, Planck's formula was quoted in several german and british sources at
the time [Kuhn 78].

\medskip\par

Rayleigh had pointed out in June, 1900, that classical theory would predict
$U = kT$ for the oscillator energy (for it $\langle E_{\rm
kin}\rangle = \langle
E_{\rm pot}\rangle$, each with $kT/2$),  from the equipartition theorem, and by
1905 he, Einstein and Jeans were firm that the exact prediction of classical
physics for the Kirchhoff function had to be
\begin{equation}
u(\nu, T) =  8 \pi (\nu^2/c^3) kT,
\end{equation}
absurd, as no maximum in $\nu$ implied divergence for the total
radiation at any
temperature, the (later) UV catastrophe of Ehrenfest. Formula (13) became
to be known as the Rayleigh-Jeans law (R-J). Notice the two former constants
conspire to leave only one, $k$: classical  physics was absolutely unable to
produce the maximum of the radiation formula, because this requires two
{\sl disentangled} constants! The same conclusion (13) was reached by
Lorentz in
1908 (Rome lecture), when the electrons, by then secure componentes of
elementary matter, took the place of the imaginary oscillators. It is not
sufficiently emphasized that turn-of-century theoretical physics was not only
getting away from {\sl experientia} (specific heats, black body
radiation, motion
of the aether: the ``clouds'' of Kelvin), but just becoming
inconsistent; another
paradox was discovered by J. W. Gibbs, with mixtures of nearly, but not quite,
equal molecular species.

\medskip\par

By 1912 irrefutable proofs of the need for ``quantization'' were provided by
Einstein (1905), Lorentz (Wolfskehl G\"otttingen lectures, 1910),
Poincar\'e (1912
paper, [Prentis 95]), etc., see [Hermann 69]. As for Planck, he remained in an
uneasy position; for a long time he tried to find room for $h$ in the framework
of classical physics. He even thought the burden will be buried in the atomic
particles (electrons etc.) which started to proliferate around 1900. Then he
abandoned, and around 1910-12 he developed the so-called second radiation
theory. Planck was very german and conservative: he wanted the quantum to
carry the less possible damage to classical physics; also he was 42 in 1900.
One important point, though, appeared in the first edition of his {\sl
Vorlesungen} (1906) [Planck 06]. Introducing phase space, an idea he seems to
borrow from Gibbs, he concluded on the expression  $\varepsilon = h \nu $ not
from ``correspondence'' with the displacement law, but from equal-area ellipses
of the oscillator. This opened Planck's eyes to the meaning of $h$ more than
anything else, and he henceforth refers to $h$ as ``the elementary quantum of
action'' ({\sl elementares Wirkungsquantum}). More on that later.

\medskip\par

Paul EHRENFEST was a singular figure in physics in the first third of the XX
century; he was to be a critical mind for the quantum theory, and enjoyed
particular friendship with both Einstein and Bohr. He realized also, around
1905/6, that Planck had tacitly quantized the ellipses, and that that was the
real novelty. Ehrenfest also studied carefully the approach to equilibrium
(randomization) of the natural radiation, see the long discussion in [Kuhn 78,
pp. 152-169]; the treatment in [Mehra 82, Vol I, pt. 1], is much shorter. He
also worried about the counting of states, and the difference, for light
quanta (see next), between Planck's formula and the Wien limit. See also
[Navarro 03].

\medskip\par

The article of Albert EINSTEIN in 1905 dealing with the light quantum
hypothesis is so well known that we shall be brief. As regards the black body
radiation, the main point is that the Wien limit of the Planck's formula is
recovered supposing that radiation is composed of {\sl corpuscular} grains of
energy
$\varepsilon = h \nu $, because the computable entropy function coincides with
that obtained by Planck to justify, in  early 1900, Wien's formula. Of course,
there are more reasons to introduce the {\sl Lichtquanta}, as Einstein
called them;
the photoelectric effect is the best known, but the original (and
best, we think)
reason is to cure the asymmetry between matter and radiation if the later is
continuous (Einstein points here clearly to the UV catastrophe); this
magistral idea of Einstein, of denouncing unphysical asymetries in classical
physics was already used, equally succesfully, in the starting paragraphs of
the special relativity paper, also in his {\sl Annus Mirabilis} of
1905: the coil
vs. magnet motion {\sl Irrlehre}, as pre-relativity theory predicted different
results according which is moving.

\medskip\par

So Einstein's light quanta reproduce the low-density (Wien)  limit of Planck's
formula; otherwise Einstein was very careful not to endow the (future) photons
with too much ontology; in  particular, until 1917 no momentum was asigned ($p
= h\nu/c$), although momentum fluctuations did appear in a 1909 paper. Question
arises (today), if light is really corpuscular, why he (E.) did not obtain the
correct (Planck) formula? There is some discussion for this in the literature,
old and modern; we feel that Eistein was unable to do it: he could not, in
early 1905, obtain the correct radiation formula from the bare ``photon''
hypothesis, so he published just the Wien ``approximation'' \footnote{I thank
L. Navarro (Barcelona) for a discussion of this point}. We shall use the term
{\sl photons} (G.N. Lewis, 1.926) as a commodity for light quanta.

\medskip\par

There are two different arguments, given today, for that oddity: first, the
``duality'' wave-particle. Namely Einstein himself proved in 1909, that the
fluctuation formula for the energy

\begin{equation}
\Delta E^2 \equiv \left\langle (E - \langle E\rangle)^2\right\rangle
\end{equation}
applied to Planck's formula produces straightforwardly
\begin{equation}
  \Delta E^2 = h \nu \langle E\rangle  + \frac{c^3}{8\pi \nu^2}
\langle E\rangle^2
\end{equation}
(per unit volume and frequency range), where $\langle E\rangle  = U$
is the mean
oscillator energy.

\medskip\par

This is a very remarkable formula. The split in it is identical  as the split
$R = \alpha U + \beta U^2$, the starting point of Planck's phenomenological
deduction of his formula, Cfr. (10) above. This has been noticed already, see
e.g. [Hermann 69, p. 59]. Therefore, the linear part in (15) corresponds to the
Wien (corpuscular!) limit, the quadratic part to the R--J (wavelike!?) limit.

\medskip\par

Einstein himself adopted that interpretation, and this is one of the reasons,
we reckon, for the widespread propagation of the idea of the particle-wave
duality; Einstein stated explicitely that in view of this, he expected ``the
next development in theoretical physics will provide a theory of light to be
interpreted as a kind of fusion of the wave and emission [corpuscular]
theories'' [Pais 82, 21a]. We believe this is a misleading idea, and shall
comment upon it later.

\medskip\par

The second argument contradicts the former. It is this: the  correct Planck's
formula can be obtained, from the purely corpuscular point of view, {\sl taking
in account the photon indistinguishability}. This was done by Einstein in 1906,
by Debye in 1910 and by Bose in 1924, without mentioning, none of the three,
indistinguishability! The effect is dissimulated by a different, peculiar
counting. For example, taking the light as composed of ``molecules'' of energy
$n h \nu$ as separated entities, one arrives easily (see Appendix) to
the correct
radiation formula; this is the Einstein derivation; in this (correct!)
approach, the split in (15) has to be interpreted just contrary to usual: for
low densities, the correlation efect, which is quantal, as it comes from
identical particles, does not show up; so the Wien limit of the radiation
formula is ``classical'', for corpuscles, as uncorrelated massless particles,
whereas the R-J limit is the pure quantum part!! Please notice this is
counterintuitive: one expects quantum effects to be noticeable for low T,
contrary to our case here; this is again an effect of zero mass of the
photons. As for Debye's calculation, it is formally the same as Einsteins's,
but the prefactor $8\pi \nu^2/c^3$ is taken as the number of resonance modes in
the cavity, plus the Planck energy quantization: so Debye gets rid  of the
oscillator altogether; for the derivation see [Born 46, VII-1]. The derivation
by Bose (1924) also disposes of the fictitious oscillators.

\medskip\par

{\sl The identity of particles provides a nonclassical
interdependence of photons
which mimics the ondulatory properties}. But this was never understood by
Planck,  Einstein Debye or Bose (although Plancks gets close to justify a good
counting with identical particles; see [Rosenfeld 36]); when Einstein
introduced the Bose-Einstein statistics in 1925 he notes the interdependence,
but just says it is mysterious. The thing only became clear as a corollary of
Heisenberg's and Dirac's first treatment of identity of particles in the new
quantum mechanics, in 1926; it is a beautiful example of the power of the
healthy positivistic attitude in science: interchange of identical particles
is unobservable, {\sl therefore} the theory should abide by it; in fact,
Leibnitz
already thought along these lines.

\medskip\par

It follows that the wave-particle duality interpretation must be  false!
Indeed, this much is stated in [Bach 89]. He claims that Einstein made a
mistake in his ``dual'' (1909) interpretation of (15). Bach  concludes that two
terms contributing independently to the fluctuation formula is the {\sl wrong}
inference from probability theory, and that Einstein interpretation is false;
but it will take us too afar to delve in this, to which we want to come back
in the future. So we content ourselves here to add some comments on the
attempts to understand the Einstein-Wien (independent) photons vs. the
Einstein-Planck (interdependent) photons in the old quantum theory. This point
was considered by Ehrenfest, Natanson, Wolfke, Krutkow and de Broglie, in
1910-23; in particular the latter aimed to a particle description of
interference, a seemingly impossible task, but understood today through
Feynman's path integral formulation of quantum mechanics. A thorough study has
been published recently [Perez-Canals 02] and we refer to it. The historical
controversy is well told in [Jammer 66, pp. 50-52], in [Mehra 82, I-2 p. 559]
and in [Whittaker 10, pp. 102-104]; the later is one of the few sources, as far
as we are aware, to emphasize explicitely that a purely corpuscular theory of
light {\sl can} produce Planck's formula. This is with hindsight, of course!

\medskip\par

Among the other contributors to quantum theory up to 1912,   we just mention
two. A. Haas was the first (1910) on thinking of a conection between the
quantum, and atomic discreteness: he wanted to explain the quantum of action
in terms of the atoms;  he found a relation between the radius of the Thomson
atom and the quantum of action; for a delightful exposition see [Hermann 69,
pp. 91 ff.]; Haas' theory is a direct antecedent of Bohr's atom, see [Heilbron
69]. A. SOMMERFELD rightly pointed out [Hermann Ch. 6], in the Karlsruhe and
Solvay 1911 conferences, that it was the other way around: it is the
existence, stability and excitation of atoms which should be understood in
terms of Planck's constant. Other important point was realized by several
people (Planck, Einstein, \ldots ): the dimensions of the quantum of action $h$
is the same as that of $e^2/c$; so some attemps were made to relate the quantum
of action $h$ to the ``quantum'' of electricity $e$. Today, the mystery of the
value of the fine-structure constant $ \alpha = e^2/hc \approx (137)^{-1}$ is
still with us \ldots .

\medskip\par

{\bf 6.-- A quick look at Planck's second theory 1910-1912 is in time
now, and appropiate}. It seems that Planck hit upon it after reading Lorentz's
G\"ottingen lecture, accepting the quanta (quoted in [Klein 66]). Planck has
accepted discontinuity, but wants to put it where it makes less harm, another
confirmation of his conservative character. He prefers to quantize the
oscillator, and leave the radiation, as in the wave theory, continuous; he
cannot be blamed for the later: except Einstein and Stark, nobody accepted the
{\sl Lichtquanta} yet. Quantization takes place in phase space now, and as such
this is a forerunner of the subsequent quantization rules of the
Bohr--Sommerfeld
atom. In fact, the formula that Planck writes (1911)
\begin{equation}
\Int\Int dp \,dq = nh
\end{equation}
has a deep invariant meaning, as we know today, through the symplectic
approach to mechanics, because $dp \wedge dq$ is just the symplectic
2--form. More
important still was the ``average'' over the oscillator energies, from which
Planck concludes that
\begin{equation}
  E = (E_n + E_{n+1})/2 = (n + 1/2) h \nu
\end{equation}
and the half-quantum makes his first appearance in physics, not to
  be rediscovered  until 1925 \ldots .

\medskip\par

This second theory of Planck is not very interesting today; for an original
interpretation (based in putting discontinuity at the start) see [Kuhn 78, pt.
3].

\medskip\par

For lack of space we refrain to gloss over the important role
  Planck and Einstein played in the development of the third principle of
thermodynamics (Nerst) and in the zero-point energy issue.

\medskip\par

\medskip\par

{\bf 7.-- The mood was not very inclined towards historical studies of  the
quanta until 1960}, and here only a few papers are examined. Some contributions
of L. Rosenfeld, a close collaborator of Bohr, (and, according to
Pauli, {\sl der
Chorknabe des Papstes}) are worth commenting. In his [Rosenfeld 36] paper he
recounts carefully Planck's achievement. One nice point he emphasizes is the
transition of the variables density $u$ and entropy $s$, pertaining to the
radiation, to the corresponding $U$ and $S$ belonging to the oscillator; in
particular, from $\partial s/\partial u = \partial S/\partial U$, and
the $s(u)$
relation
  from Wien's law, Planck concludes the form (8) for $S(U)$, which then uses to
justify (9). Is the formula unique? Rosenfeld recalls the important relation
\begin{equation}
\Frac{d \Sigma}{dt} = \Frac{3}{5}\Frac{ d^ 2S}{dU^2} \,
   \Frac{dU}{dt}\, \Delta U
\end{equation}
for the time variation of the total entropy $\Sigma$; as $(dU/dt)\, \Delta
U$ is
negative, increasing entropy means that
\begin{equation}
R \equiv - \left(\Frac{d^2S}{dU^2} \right)^{}  \ge 0
\end{equation}
which is satisfied by $ R = \alpha U$, $\alpha > 0$, but by many
others as well.

\medskip\par

Wien's law is not unique. Rosenfeld, probably rightly, says that  Planck
stumbled in the form $R = \alpha U + \beta U^2 $ not only because the
second term
would reproduce the experimental finding $u\propto T $ at hight $T$ (low $\nu$)
that Rubens reported to Planck, but also because it still gives the
``logarithmic'' aspect to the function $S=S(U)$ that pleased Planck because
similarity with Boltzmann's formulas. Other quantum papers by Rosenfeld in the
book referred in [Rosenfeld 36] are worth reading.

\medskip\par

M. J. Klein wrote an important paper in 1961 [Klein 61; see also
  Klein 66]; in fact, one can consider Klein the forerunner of the critical
historians of  the quantum theory, anteceding Kuhn and Jammer. Here, however,
we shall give him just a cursory examination. He deals with two specific
points. (i) Did Planck know  in October 1900, about the Rayleigh-Jeans
formula, that Rayleigh had published in June, 1900? Answer: probably yes, but
he did not pay attention; Klein gives plausible reasons for both statements;
we tend to agree. By the way, the polemic about previous knowledge of a
clearly antecedent paper repeats itself in Einstein with respect to the
Michelson-Morley experiment, and with Bohr and the Balmer formula. (ii) In
what ways did Planck depart from Boltzmann's methods in his statistical
calculation? Klein signals a few; Planck calculated the complexions
(microstates) of a single macrostate, whereas Boltzmann compared the relative
weights of different states, looking for maximalization. In fact, it seems
that Planck also in his 14-XII-1900 derivation was again working backwards
(because he knew the correct answer), a suggesion from [Rosenfeld 36] that
Klein accepts. And also the peculiar counting method of Planck is at odds with
the distinguishable entities Boltzmann had considered, a point that Ehrenfest
forcefully expressed in 1911 and later (compare our comments above).

\medskip\par

As his final point, Klein comments on the little interest in  Planck's theory
up to 1905, and adds an extra reason to the exposed above: many continental
theorists, up to 1910, were recalcitrant antiatomists, defending the
``energeticism'' point of view; among them Ostwald, Mach and Duhem; the battle
Boltzmann fought against them (and that perhaps costed him his life) is well
documented.

\medskip\par

The history of photon statistics is told in [ter Haar 69]. The
  paper makes good historical points (such as the diverse names attached to
$k$,
the role of Ehrenfest, etc.), but the discussion of the photon statistics
issue is poor (e.g., ignores the Natanson  vs. Krutkow controversy, and also
the issue of identity vs. indistinguishability, etc.).

\medskip\par

The contribution of Rest Jost in the Einstein Centenary [Jost 79]  is the best
source for the atomic controversy between Boltzmann and Planck, with Mach as
inspiration and Einstein as spectator; the paper is nearly philosophical, and
there is no substitute for reading it. Planck started to reject atoms because
with them, the proof of the second law was not absolute, and ended in just the
opposite: irreversibility leads to the atoms!

\medskip\par

Stephen G. Brush is a recognized science historian; in his paper  [Brush 02]
on cautious revolutionaries, he, of course, looks at Planck. He reminds us of
Planck's principle: the new scientific ideas triumph on the long run, because
its opponents die, not because they became convinced. Brush also takes part
with Kuhn's points of view, and provides powerful arguments (not wholly
convincing, we think; for example, Brush insists in Planck's rejection of the
photons; but, as we argued, everybody did \ldots ).

\bigskip\par
\medskip\par

{\bf 8.-- In referring to the papers appearing for the centenary, the
  one by G. Parisi [Parisi 01] is interesting}; he remarks about lack of simple
proof of Wien's displacement formula (2); we hope our deduction in the
Appendix is simple. Also, he notices that Planck had introduced transition
probabilities in this second theory, anteceding Einstein for five years; he
also points out the ``mistake'' made by Bose in his derivation. The best part,
we think, is in the final: we learn that Jeans was nearly correct in supposing
that thermodynamical equilibrium had not been obtained (it requires a very
long time), and also that the correct behaviour of the equilibrium time for
small coupling is not known (even today!).

\medskip\par

[Rechenberg 00] repeats his plea [Mehra 82] against Kuhn on
what discontuity did Planck find; we commented this point above; we disagree
that classical physics was not completed until 1905, if only because special
relativity meant such a enormous conceptual break (e.g., relative time); it is
nice, though, that Rechenberg remind us that Planck saw the action is a
relativistic invariant.

\medskip\par

[Studart 01] is correct and fairly complete. The combinatorial
deduction of the radiation formula is very detailed; it is interesting the
suggestion of Planck as a ``sonambule'' of science in the sense of A. Koestler.

\medskip\par

[Sanchez-Ron 00] comments Planck's achievement on the
  conmemorative issue of the ``Revista Espa\~nola de F\'{\i}sica''. He
reproduces a long
part of the letter of Planck to Wood (1931), the best testimony we have on
Planck's march towards his quantum. Another spanish author [Zamora 01] is
worth looking at for his short but accurate historical presentation. The
November 2001 issue of the Bulletin of the Mexican Physical Society is also
fully devoted to the celebration [Mex 01].

\medskip\par

The triumph (radiation formula) and failure (second law) of
Planck is beautifully told in [Straumann 00].

\medskip\par

\medskip\par

{\bf 9.-- We would like to finish by launching a last look to Planck and
his oeuvre}. For the dilemmas an upright man like him faced, see [Heilbron 86].
A centennary volume with some partial reprints is [Duck 00]. Planck represents
the best of western tradition in science, an archetype which tends to
dissapear: the turgid, respected and serious german professor, coming from an
academic tradition, like Bohr, Pauli and Heisenberg; Planck was not a genius,
and he knew that. His long life represents the zenith and the fall of the
german science as no other man does; his personal tragedies would caused doom
in anyone else with less stamina. His honesty and {\sl bonhommie} were
legendary.
With Planck's death in 1947 the 25--centuries domination of European science
comes to an end, and the leadership crosses the oceans. Let us hope it will be
for good;  the new style is already different \ldots .

\medskip\par

Three final testimonies for our man seem appropiated:

\medskip\par

\hfill
\begin{minipage}[t]{14cm}
``Very few will remain in the shrine of science, if we eliminate those moved by
ambition, calculation, of whatever personal motivations; one of them will be
Max Planck''. \medskip\par
\qquad \qquad A. Einstein  in [Planck 81]
\end{minipage}

\bigskip\par
\hfill
\begin{minipage}[t]{14cm}
``The wish to preceive \ldots\ the prestablished harmony is the source of the
inexhaustible pacience and tenacity that we see in Planck as he strugles with
scientific problems, without deviation to simpler and even profitable
objetives. Colleagues attribute this attitude to an exceptional strong will
and discipline. I believe this is a complete mistake. The emotion provided by
these achievements is analogous to the religious experience or to falling in
love; the daily effort does not come from design or program, but from a sheer
and direct need''.
\medskip\par
\qquad \qquad 		A. Einstein in [Pais 82, 2a]
   \end{minipage}

\medskip\par

\bigskip\par
\hfill
\begin{minipage}[t]{14cm}
`` The profound purity of his creation, the clarity, depth and and deceptive
simplicity of his thinking, and the lifelong nobility of his character and his
uncompromised principles are his everlasting monument''.
            \medskip\par
\qquad \qquad   E. C. G. Sudarshan in [Duck 00]
\end{minipage}
\medskip\par

  \newpage
\centerline{\bf  APPENDIX  }
\bigskip\par

\noindent{\bf A. Relation between intensity $K$ and density $u$:}
\medskip\par
In general
current  is density times velocity, $\Vec{j} = u \Vec{v}$; but here
intensity $K$
is current flux through the unit sphere, and $v = c$; hence
\[
u(\nu, T)= 4 \pi K(\nu ,T)/c.
\]
\bigskip\par

\noindent{\bf B.	The derivation of the Stefan-Boltzmann law:}
\medskip\par
If
heat $\delta Q$ is added to the cavity, the increase has a part due to the
internal energy $\delta U$ and a part due to the work performed by expansion $p
\, \delta V$; this is the first principle:
\[
\delta Q = \delta U + p \delta V.
\]
Now  $U=uV$, where $u$ is the density of energy, function {\sl only}
of $T$, the
absolute temperature; for electromagnetic waves, the radiation pressure is $p =
u/3$, as deduced by Boltzmann from Maxwell equations (but not checked
experimentally until 1902 (Lebedev)). Now $\delta Q/T = dS$, where $S$, the
entropy, is a local function (hence $dS$ is an exact form); therefore, $S =
S(T,V)$ and
\[
dS = \Frac{V}{T}  \Frac{\partial u}{\partial T} dT  + \Frac{4}{3}
      \Frac{u}{T} d V.
\]

	Exactness implies
  \[
  \Frac{\partial((V/T)u'(T))}{\partial V}  =
   \Frac{\partial((4/3)(u/T))}{\partial T} = \Frac{u'}{T},
  \]
  or
  \[
  \Frac{u'}{T} =  \Frac{4}{3}\left( \Frac{u'}{T} -  \Frac{u}{T^2}\right),
  \quad   \mbox{or}   \quad    u(T) = \sigma T^4,
  \]
which is the Stefan-Boltzmann law.

\medskip\par

\bigskip\par

\noindent{\bf C. Derivation of the Wien's displacement law}
\medskip\par
The usual derivation of the displacement law  (Wien) is
cumbersome: [Parisi 01] complaints. The simplest is the following: for
adiabatic changes  $\delta Q = 0$, hence
\[
  (4/3)u dV + V du = 0 ,
\]
or
\[
V u^{3/4} = \mbox{const.} \quad \mbox{or} \quad
  V T^3 = \mbox{const.}, \quad \mbox{or}  \quad
\lambda T = \mbox{const.},
\]
because the increase in $V^{1/3}$ is linear with the increase in $\lambda$
(Doppler effect). This is strictly speaking, Wien's displacement
law; hence $T$ enters in $u(\nu,T)$ only in the form $T/\nu$.

\medskip\par

Now from the Stefan-Boltzmann result,
\[
\Int_0^\infty u(\nu, \nu/T) d\nu = \sigma T^4 = T \Int_0^\infty u(xT,x) dx ,
\]
which implies, for $u$ positive,
\[
\Frac{\partial^4}{\partial y^4} u \left(y \equiv xT=\nu, x\right) =0,
\]
or $ u(y, x)= y^3 \phi (x)$ neglecting lower powers from the S-B result. So
\[
			u(y, x) = u(\nu, \nu /T) = \nu^3 \phi(\nu/T),
	\]
which is the usual form of Wien's law.

\bigskip\par

\noindent{\bf D. The forced damped oscillator. }
\medskip\par
The equation (6) in the $x$-axis
for  a single mode $\Omega$ of the e.m. field is
\[
		 m \ddot {x} + \gamma \dot {x} + kx = e {\cal E}
\cos(\Omega t);
\]
written in the operator form
$(D-a)(D-b)x=(e{\cal E} /m)\cos(\Omega t)$, where $a$ and $b$ are
roots of  $y^2 +(\gamma /m)y + \omega^2=0$, and $D = d/dx$, the equation admits
as solution, as $D-a = \exp(at)\,D \,\exp(-at)$ etc.,
\[
			 x = \mbox{transient} + C \cos(\Omega t - \theta)
\]
where $C$ and $\theta$ can be calculated at once; the transient decays with the
time constant  $\gamma /m $. Now the {\sl instantaneous} oscillator energy $E =
E(\omega, \Omega)$ is $k x^2_{\max}/2$,  $x_{\max}=C$, and hence
\[
			E = \Frac{(e^2 {\cal E}^2/2m)}{( \Omega-\omega)^2 +
(\gamma/m)^2}.
\]
\
	Notice how the damping avoids the blow-up for $\omega \to \Omega$.
\medskip\par
Now the energy density for the e.m. field $u = u(\Omega)$ is
$({\cal E}^2 + {\cal H}^2 )/8\pi$, and it is supposed isotropic, i.e.
$u = 6 {\cal E}^2/8\pi$;
the {\sl equilibrium} oscillator energy  $U(\omega)$ is obtained by integration
on $\Omega$. Only the resonance frequency contributes, as $\gamma$ is
very small.
The final result is the relation (restoring $\nu = \omega /2 \pi$)
\[
			u(\nu, T) = 8\pi \nu^ 2/c^3\, U(\nu, T),
\]
(where $T$ is there just for the ride) written in (7).

\bigskip\par

\noindent{\bf E. Planck's Derivation of Wien's and Planck's formula.}
\medskip\par
From
\[
			R \equiv - \left( \Frac{\partial^2 S}{\partial
U^2} \right)^{-1} = \alpha U
\]
  we get
  \[
\Frac{\partial S}{\partial U} = -\Frac{1}{\alpha} \log U + b,
\]
hence,
\[
dS = \left(\Frac{- \log U}{\alpha} + b\right) \, dU \, \equiv \Frac{dU}{T},
\]
or
\[
		U = K \exp(- \alpha/T), \qquad  \mbox{ Wien's law in }  \,\,T.
\]

	Now from
  \[
	R \equiv - \left( \Frac{\partial^2 S}{\partial U^2} \right)^{-1}
          = \alpha U + \beta U^2,
\]
identical calculation gives (the integration constant is fixed)
\[
					 U = K/[\exp(+\alpha /T) -1],
\qquad  \mbox{ Planck's law}.
\]
Wien's form $\Longrightarrow  K, a \propto \nu$.

\medskip\par
The full entropy obtains after another integration; write first the full
actual equation
\[
			U(\nu, T) = h \nu /[ \exp(h\nu /kT) -1]
\qquad  \mbox{ Pl.}
\]
and another integration (no constant) gives
\[
		S(U) = k[ \log(1+U/h \nu)^{1+U/h\nu}  -
\log(U/h\nu)^{U/h\nu} ],
\]
which is of the form $A = k \log W$, and gave to Planck the idea of a
combinatorial approach.

\bigskip\par

\noindent{\bf F. Einstein's derivation of Planck's formula.}
\medskip\par
If the oscillator
is  quantized, $E(n) = nh \nu$, the mean energy is
\[
\langle E \rangle = U = \Frac{\partial \log Z}{\partial \beta},
\]
with
\[
Z = Tr \exp(-\beta H)= \Sum_0^\infty x^n,\qquad \mbox{and} \qquad
x = \exp(-\beta h \nu);
\]
so $Z(\beta) = (1-x)^{-1}$,  and
\[
U =\Frac{h \nu}{\exp(h \nu /kT) - 1}.
   \]
\medskip\par

\newpage
\section*{BIBLIOGRAPHICAL   NOTES}

\medskip\par

The literature on the black body radiation is overwhelming. There is no
question here of quoting the earliest primary sources, as they are difficult
to consult, and are called for in most good modern treatises; so here we cite
the most important secondary sources.

\medskip\par

The early history of the radiation formula is best told in [Kangro 76]. A
detailed information on the experimental situation is well described in
[Sanchez-Ron 01]. The contribution of Planck is told in many sources; the best
simplest is perhaps [Hermann 69], Ch. 1; the encyclopedic work of Mehra and
Rechenberg [Mehra 82] deals with Planck in Vol 1, Part 1. The monography by
[Kuhn 78] is unsurpassed on the work of Planck and its antecedents; Kuhn's
interpretation is however somewhat controversial, see our main text. A
centenary book with some reprints is [Duck 00]. The derivations we give on the
Appendix  are shorter that those given in many places; the most quoted ones
are in [Born 46], App. 27. The remark that Wien law leads to the inconsistency
$u$ const for $T$ large is in the undergraduate textbook [Cabrera 50], Cap. 34.
The famous {\sl Vorlesungen} of Planck are commented, reproduced and
translated in
[Planck 06]; there are some changes in the (later) english translation. The
biography of Einstein by Pais [Pais 82], VI, 18 \& 19 is worth looking at in
respect to Planck's work and antecedents.

\medskip\par

There are two old masterly expositions of the Old Quantum Theory with chapters
on {\sl W\"armestrahlung}, by [Pauli 26] and by [Rubinowicz 33]; the first
follows an
unconventional order, Planck's formula is discussed in Sect. 14, but is worth
reading (Pauli is a master of exposition); the second is more encyclopedic, as
befits to a Sommerfeld's student. There are of course some well known books
related to historical studies on the development of quantum theory: [ter Haar
67] devotes a large chapter to Planck and includes an english version of
Planck's October and December (1900) fundamental communications. [Whittaker
10] is as meticulous as always, and in his study of Planck uncontroversial.
[Hund 67] is remarkable for his emphasis on indistinguishability and
statistics, and he rightly stresses the contribution of Natanson and others.
There is the monumental japanese work [Taketani 01], with brief and accurate
calculations (Vol. II), although the style is peculiar at times. But the best
source is perhaps [Jammer 66], both in accuracy and extension and in critical
analysis.

\medskip\par

Planck's own recollections are given in his Nobel Lecture [Nobel 18] and his
Autobiography [Planck 43]. Previous to the centenary there are few articles on
Planck's discovery. We have consulted  [Rosenfeld 36], [Klein 61], [Klein 66],
[ter Haar 69] (on photon statistics). The perils of philosophers dwelling on
sheer physical issues are well illustrated in [Agassi 67]. The opposing views
of Boltzmann and (young) Planck as regards Atomism are very well expressed in
[Jost 79]. Poincar\'e last contribution (1912) is devoted to a proof of quantum
discontinuity, and is glossed in [Prentis 95]; his importance, however, is
played down by Kuhn (op. cit.) To know more about Planck directly one should
consult the very accesible Autobiography and his Lectures on Thermodynamics
[97].

\medskip\par

Finally we come to the recent works on ocassion of the centenary. As they are
commented upon on the main text, we just select some briefly here:
[Sanchez-Ron 00],
 [Parisi 01], [Mex 01], and [Studart 01]. A late addition is [Campos 04].


\end{document}